\documentclass[conference]{IEEEtran}
\ifCLASSINFOpdf
\else
\fi
\usepackage{multirow}
\usepackage{xcolor}
\usepackage{soul}
\usepackage[normalem]{ulem}
\usepackage{amsmath}
\usepackage[pdftex]{graphicx}
\usepackage{cite}
\usepackage[bookmarks=false]{hyperref}
\usepackage{algorithm}
\usepackage{algpseudocode}
\usepackage{amsthm}
\newtheorem{definition}{Definition}

\begin{document}
\title{A Survey of Potential MPI Complex Collectives: Large-Scale Mining and Analysis of HPC Applications
% \title{FP-AMG: A Reconfigurable FPGA-based Framework for Acceleration of Algebraic Multigrid Solvers
% \vspace*{-0.2truein}} 
}

\author{\IEEEauthorblockN{Pouya Haghi\IEEEauthorrefmark{1},
Ryan Marshall\IEEEauthorrefmark{2}, Po Hao Chen\IEEEauthorrefmark{1}, Anthony Skjellum\IEEEauthorrefmark{3}, Martin Herbordt\IEEEauthorrefmark{1}}
% \IEEEauthorblockA{\IEEEauthorrefmark{1}Department of Electrical and Computer Engineering, Boston University, Boston, MA, USA\\
% \IEEEauthorrefmark{2}Department of Physics, University of Science and Technology of China, Hefei, China\\
Email: \IEEEauthorrefmark{1}haghi@bu.edu, \IEEEauthorrefmark{2}rjmarshall2@ua.edu, \IEEEauthorrefmark{1}bupochen@bu.edu, \IEEEauthorrefmark{3}tony-skjellum@utc.edu, \IEEEauthorrefmark{1}herbordt@bu.edu}

\maketitle

\begin{abstract}
Offload of MPI collectives to network devices, e.g., NICs and switches, is being implemented as an effective mechanism to improve application performance by reducing inter- and intra-node communication and bypassing MPI software layers. Given the rich deployment of accelerators and programmable NICs/switches in data centers, we posit that there is an opportunity to further improve performance by extending this idea (of in-network collective processing) to a new class of more complex collectives. The most basic type of complex collective is the fusion of existing collectives. 

In previous work we have demonstrated the efficacy of this additional hardware and software support and shown that it can substantially improve the performance of certain applications. In this work we extend this approach. We seek to characterize a large number of MPI applications to determine {\it overall} applicability, both breadth and type, and so provide insight for hardware designers and MPI developers about future offload possibilities. 

Besides increasing the scope of prior surveys to include finding (potential) new MPI constructs, we also tap into new methods to extend the survey process. Prior surveys on MPI usage considered lists of applications constructed based on application developers' knowledge. The approach taken in this paper, however, is based on an automated \textit{mining} of a large collection of code sources. More specifically, the mining is accomplished by GitHub REST APIs. We use a database management system to store the results and to answer queries. Another advantage is that this approach provides support for a more complex analysis of MPI usage, which is accomplished by user queries.
\end{abstract}

\begin {IEEEkeywords}
High Performance Computing, MPI, Mining, Survey, Automation
\end{IEEEkeywords}

\maketitle
% \vspace{-0.1truein}
\section{Introduction}

Message Passing Interface (MPI) \cite{walker1996} is the \textit{de facto} standard in HPC and it has been widely used to implement portable and scalable parallel applications. It is being actively supported and developed by dozens of implementations.

MPI offers various primitives; among them collectives are integral part of MPI and they are frequently invoked in a spectrum of HPC applications \cite{zhou19}. Offloading MPI collectives to network devices (NICs and switches) is gaining much interest as an effective mechanism to improve the application performance \cite{in_nic_collective1,Haghi20a,Bayatpour2021,Graham2016,desensi2021flare,Stern18,Haghi22,Haghi23}. More specifically, in-network processing unlocks higher application performance by reducing inter- and intra-node communication and bypassing MPI software layers. As new classes of devices including programmable NICs/switches \cite{Liu2019,Bosshart2013}, Data Processing Units (DPUs) \cite{Burstein2021}, and accelerators (FPGAs, GPUs) \cite{Haghi20b,Guo22a,Guo23} are emerging in the datacenters \cite{Shahzad22,Bobda22}, we posit that there is an unrevealed opportunity to further improve the performance by extending in-network collective processing to a new class of complex collectives. 

The most basic type of complex collective is the fusion of (more than one) existing collectives; in this paper, referred to as fused collectives. This can be either back-to-back collectives or collectives with computation in between. Complex collectives can take other types; 
% for example, there might be only one collective communication but the computation is fused to the communication. More specifically, computation following the collective depends on the collective's receive buffer or the collective's send buffer depends on the computation result. We call this type as a semi-fused collectives. 
for example, we identify a Bulk Synchronous Parallel (BSP) region \cite{Cheatham1996} as a more coarse-grained type. There are many HPC and scientific applications that follow a BSP model. For instance, in stencil computation, MPI processes communicate with neighbors and perform computation; this happens for a number of iterations. 
% Iterative solvers are another example in which each process performs part of the computation and then are synchronized at the end of each superstep. 

% In previous work \cite{Haghi20}, we have shown that offloading distributed matrix multiplication (a BSP-type complex collective) offers significant performance improvement. In this work, we take a different approach; we seek

In this work, we seek to characterize a large number of MPI applications to determine overall applicability, both breadth and type. This provides a number of benefits. First, MPI complex collective characterization can provide insight for hardware designers about future offload possibilities. Second, complex collective usage statistics can inform the MPI standardization body about standardization and prioritization of new features with the greatest and smallest impact on the community. Third, MPI programming model is extended to a new set of APIs that can abstract existing routines from application programmers' perspective.

Besides expanding the scope of prior arts to include finding (potential) new MPI constructs, we take a novel direction to attain a large-scale mining and analysis. Previous surveys on MPI usage were limited to a narrow set of MPI applications or specific projects \cite{Sultana2021, Chunduri18, Bernholdt2020, VanderWiel1997}. The approach taken in this paper, however, is based on an automated \textit{mining} of a large collection of MPI code sources hosted in GitHub. More specifically, the mining is accomplished by searching the GitHub universe using REST APIs. 
Also, prior approaches fall short to provide a large-scale complex analysis. To tackle this challenge, we use a database management system to store the information and answer the queries. Another advantage of our approach is that it provides support for a more complex analysis of MPI usage. To the best of our knowledge, this is the first work that searches a large collection of MPI code sources existing in GitHub and uses a database management system to store and analyze the MPI usage for new collectives.

% In this paper, we propose a pipeline to automate mining and analysis of large collection of MPI code sources. To the best of our knowledge, this is the first work that searches the whole GitHub universe and uses a database and a query manager to store and analyze the MPI usage for new collectives.

% Describing the pipeline:
% In the first stage of pipeline, we search GitHub universe for repositories that contain MPI collectives using GitHub REST APIs. After partitioning and cloning the resulting corpus into a disk, we analyze the code and report the MPI usage for each repository. Subsequently, the meta-information generated by the previous stage is populated in a database. Using a database management system enables us to store the information efficiently and answer the queries from the next stage. Another advantage is that this approach provides support for a more complex analysis of MPI usage, which is accomplished by user queries. Finally, in the last stage of the pipeline, an elastic query manager is able to take user input (\textit{i.e.} lists of MPI collectives to search from), generate the corresponding query to the database, and report the MPI complex collective usage.

In this work, we focus on the first type of complex collectives (fused collective) as the analysis and exploration of all of complex collectives types is beyond the scope of this paper. But our approach remains as the foundation to explore other (possible) new complex collectives and it is flexible enough to generalize to more in-depth analysis. According to our experimental results, our finding is that there is a large number of complex collective instances over the dataset we generated from open-source repositories covering a broad range of scientific domains with different sizes and level of complexity. The finding proves that there is a great opportunity for fused collectives to be further explored by MPI developers, hardware designers, and application programmers.

Our contributions are:
\begin{itemize}
    \item We generate a dataset over two hundred distinct repositories. To our best knowledge, this is the largest curation of repositories
    % with MPI collectives.
    used for MPI usage surveys.
    \item We provide a scalable framework, called MPI-Recon, for 
    % collecting more data for future studies.
    searching open-source GitHub repositories and collecting them in a database to facilitate complex analysis.
    \item We introduce the classifications of complex collectives and analyze their usages.
\end{itemize}
\section{Complex Collectives}
% HPC applications are comprised of communication routines and computation blocks. 
The idea of complex collective is to extend existing collectives to a new abstract form encompassing both MPI communication primitives and user-defined computation. Let us first define these compund collectives.

\subsection{Definition}
The most basic type of complex collective is the fusion of (more than one) existing collective communication routines. We refer to them as fused collectives. This can be either back-to-back collectives or collectives with computation blocks in between. Complex collectives can take other types; for example, there might be only one collective communication but the computation is fused to the communication. More specifically, computation following the collective depends on the collective's receive buffer or the collective's send buffer depends on the computation result. We refer to this type as semi-fused collectives. A more coarse-grained complex collective is a Bulk Synchronous Parallel (BSP) region. For example, the critical part of stencil computations and iterative solvers are BSPs. For the rest of the paper, we consider fused collectives as complex collectives and we provide a thorough analysis on their MPI usage. 

\subsection{Motivation}
\begin{figure}
  \includegraphics[width=\linewidth]{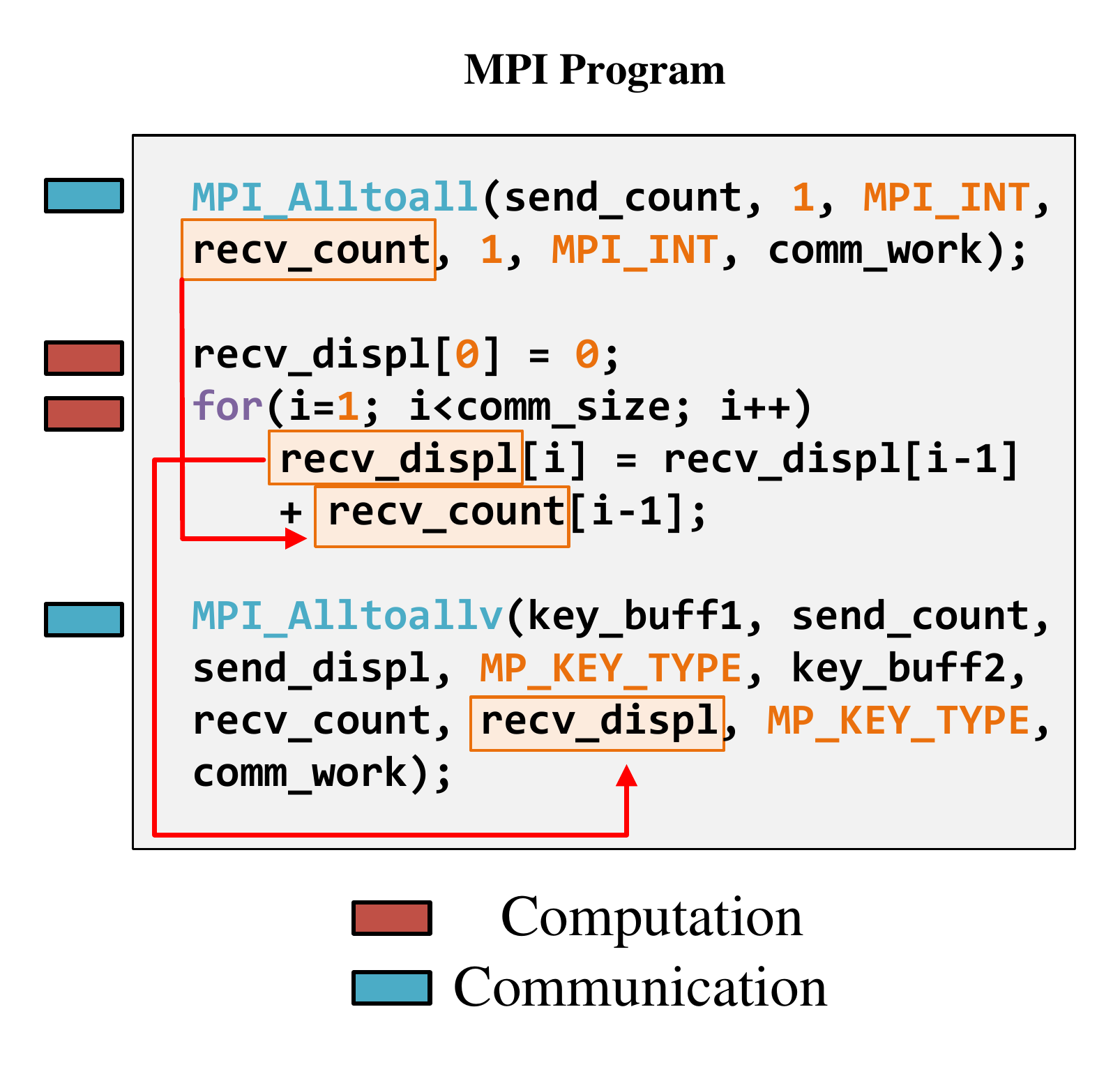}
  \vspace{-0.3truein}
  \caption{Example of a fused collective: (\texttt{Alltoall} fused with \texttt{Alltoallv} in the \textsc{IS} application from the NAS parallel benchmark.}
  \label{fig:motiv}
  \vspace{-0.1truein}
\end{figure}

We now give an example to illustrate the idea of complex collectives. Fig.~\ref{fig:motiv} (a) shows an example of a fused collective (first type) for NAS parallel benchmark \footnote{\url{https: //www.nas.nasa.gov/ publications/ npb.html\#url}}
(\textsc{IS} benchmark). Communication (marked with blue rectangles) and computation (marked with red rectangles, referred as \textit{op}) are chained together. 
% the receive buffer in the \texttt{MPI\_Allreduce} (\textit{stream\_in} array in the figure) is used in the computation and the array that \textit{op} generates (\textit{stream\_out}) is used as the send buffer by \texttt{MPI\_Alltoall}. 
The computation in \textit{op} depends on the receive buffer in the \texttt{MPI\_Alltoall} (\textit{recv\_count}); and the displacement array in the \texttt{MPI\_Alltoallv} depends on the result of computation in \textit{op} (\textit{recv\_displ}).
By offloading such a complex collective to network devices (\textit{e.g.,} switches), \textit{op} is processed in the network instead of the traditional approach: receiving the \textit{recv\_count} from the network back to the nodes, performing the \textit{op} computation in the nodes, and sending \textit{recv\_displ} to the network. This saves sending and receiving data back and forth considerably. In other words, we seek to extend this communication optimization (hop saving) from a single collective to more than one collective.

For more generic examples, there might be cases where the \textit{op} part depends on (produces) external arrays other than the arguments of preceding (following) collective. Hence, we refer to \textit{inbound} as the collection of arrays that are input to the \textit{op} and do not depend on the preceding collective(s); and similarly, we refer to \textit{outbound} as the collection of arrays that are outputs of \textit{op} and the following collective(s) do not depend on them. Clearly, inbound, input arguments of the collectives, and the target instructions in the \textit{op} (according to the architecture of programmable network devices) are sent to the network. On the other hand, outbound and output arguments of collectives are received from the network to the nodes. 

% \begin{figure}[htbp]
% \vspace{-0.2truein}
% \centering
% \includegraphics[width=1.0\linewidth]{motive_example2.png}
% \vspace*{-0.35truein}
% \caption{(a) Example of a fused collective (\texttt{Allreduce\_op\_alltoall} in the \textsc{IS} application from the NAS parallel benchmark \cite{NAS}) and (b) New translated API.}
% \label{fig:motiv}
% %\vspace{-0.1truein}
% \end{figure}

\section{Proposed Framework}
\begin{figure*}
  \includegraphics[width=0.95\linewidth]{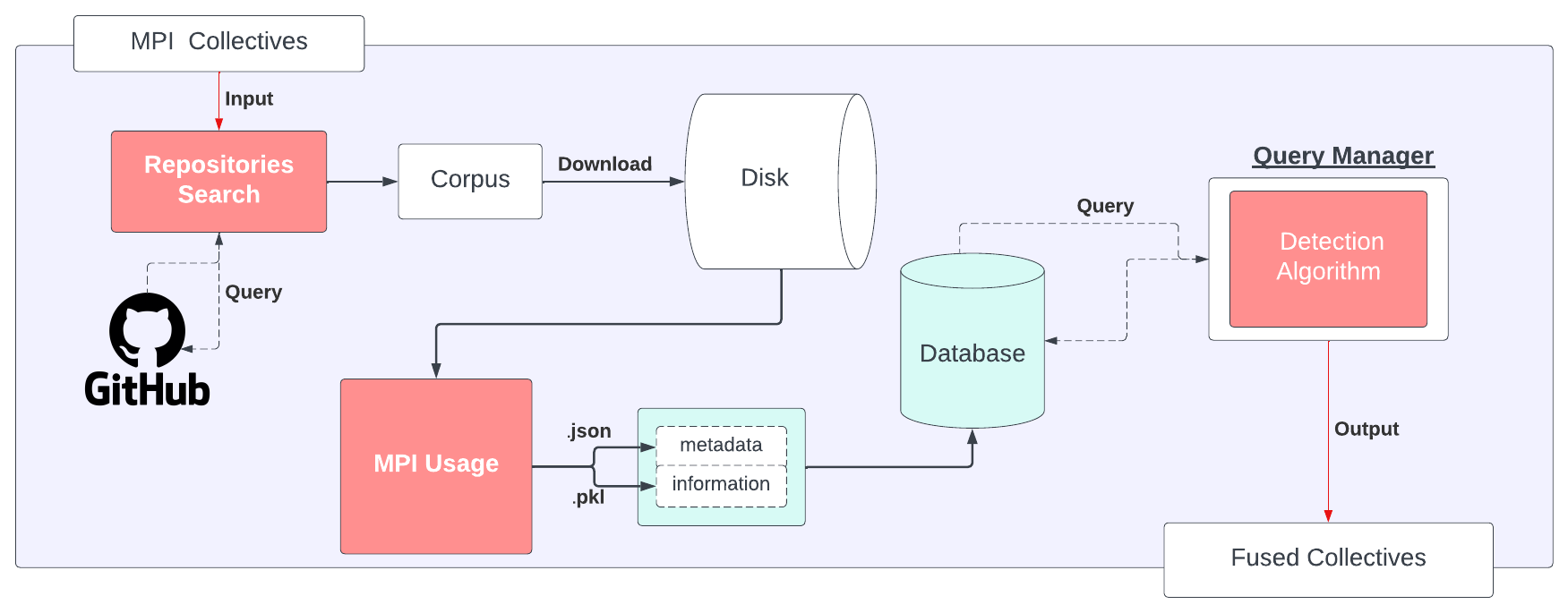}
  \caption{MPI-Recon Pipeline}
  \label{fig:mpirecon}
\end{figure*}

We describe the following setup used for our analysis. The implemented pipelines follow the depiction in Figure \ref{fig:mpirecon} and is made publicly available
\footnote{\url{https://github.com/rmarshall42/mpi-recon}}.

\subsection{MPI-Recon Pipeline}
We follow the heuristic that complex collectives exist in many classes of applications varying in sizes and structures. Our study targets open-sourced projects per the plethora of samples that are accessible.
\subsubsection{Repositories Search}
\label{sec:repo}
Previous studies \cite{Laguna2019} had a sample size up to more than one hundred distinct MPI programs. Our implementation provides a scalable method of probing for more collections via the GitHub REST APIs. We use PyGitHub \footnote{\url{https://pygithub.readthedocs.io/en/latest/introduction.html}} for this purpose.
We support searches in different programming languages. However, since the MPI Standard adopts C/C++ and Fortran as its official languages, we narrow our analysis to these. 

Our pipeline begins with the supply of MPI collectives as keywords to crawl the search space. We selected the following set in our study: "Allgather", "Allreduce", "Alltoallv", "Barrier", "Gather", "Gatherv", "Reduce", "Scatter", and "Scatterv" as they are commonly used. 
\subsubsection{Corpus}
We generate a corpus of links to the found repositories. In our study, we curated a list of over two hundred public GitHub projects. Many of the scientific applications are large in size, possibly on the scale of millions of lines in code from a relatively small sample of repositories. To avoid excessive storage space, we partition the corpus into smaller subsets and iteratively remove the downloaded content after successful extraction of the information.

\subsubsection{MPI-Usage}
We leverage the tool, \textit{MPI-Usage}, made available from previous study by Laguna et al. \cite{Laguna2019}. It recursively searches the local file directory and outputs statistics of the MPI routines in JSON format. Optionally, users may enable verbose mode to locate the files of the specific MPI calls. We processed the detailed output and saved it in standard binary format.
\subsubsection{Database}
We maintain the results from the previous stage in a SQL Database. Table \ref{tab:metadata} shows the attribute we have generated from our dataset. Naturally, duplication may arise in the corpus. We characterize them into identical code repositories or different versions of the same project. For example, MVAPICH2 vs. MVAPICH. In this case, we record both because the later version may introduce more advanced usage of MPI.

\begin{table*}
\centering
\caption{Database Attributes}
\label{tab:metadata}
\begin{tabular}{|ccccccl|}
\hline
\multicolumn{7}{|c|}{Metadata}                                                                                                                                                                                                     \\ \hline
\multicolumn{1}{|c|}{Repo ID}       & \multicolumn{1}{c|}{Owner}      & \multicolumn{1}{c|}{Filename}     & \multicolumn{1}{c|}{Revision ID} & \multicolumn{1}{c|}{Clone URL} & \multicolumn{1}{c|}{Retrieval Date} & OpenMP Lines \\ \hline
\multicolumn{1}{|c|}{OpenACC Lines} & \multicolumn{1}{c|}{CUDA Lines} & \multicolumn{1}{c|}{OpenCL Lines} & \multicolumn{1}{c|}{C Lines}     & \multicolumn{1}{c|}{CPP Lines} & \multicolumn{1}{c|}{Fortran Lines}  & Total Lines  \\ \hline
\multicolumn{7}{|c|}{Collectives}                                                                                                                                                                                                  \\ \hline
\multicolumn{2}{|c|}{Filename}                                        & \multicolumn{2}{c|}{Collective Call}                                 & \multicolumn{3}{c|}{Line Number}                                                    \\ \hline
\end{tabular}
\end{table*}

% \begin{table*}[h]
%   \caption{Database Attributes}
%   \label{tab:metadata}
%   \begin{tabular}{cclllll}
%   Metadata\\
%     \toprule \toprule
%     Repo ID & Owner  & Filename & Revision ID & Clone URL & Retrieval Date & OpenMP Lines & OpenACC Lines & CUDA Lines & OpenCL Lines & C Lines & CPP Lines & Fortran Lines & Total Lines \\
% \\
%   Collectives\\
%   \toprule
%     \toprule
%     Filename & Collective Call & Line Number
%     \bottomrule
%   \end{tabular}
% \end{table*}
\subsubsection{Query Manager} The last stage of the pipeline is an interface with the database. Our implementation supports a generic search method by sampling pairs of collective calls. This stage takes user input (\textit{i.e.} lists of MPI collectives to search from), generates the corresponding query to the database, and reports the MPI complex collective usage. The framework allows users to implement custom detection algorithm to look for undiscovered patterns.

\section{Experimental Results}
We run MPI-Recon pipeline with collectives mentioned in Section \ref{sec:repo} and the database is populated with the resulting corpus from repository search. We note that more than 200 repositories are stored in MPI-Recon database. Table \ref{tab:db} shows the total number of occurrences of each MPI collectives. \texttt{MPI\_Barrier} \texttt{MPI\_Bcast}, and \texttt{MPI\_Allreduce} have the highest number of occurrences.

\begin{table}
\caption{Database statistics: total number of occurrences of each MPI collectives in MPI-Recon database.}
\label{tab:db}
% \hline
\begin{tabular}{|c|c|c|c|}
\hline
\begin{tabular}[c]{@{}c@{}}Collective\\ Name\end{tabular} & \begin{tabular}[c]{@{}c@{}}\# Occurrences\\ in Database\end{tabular} & \begin{tabular}[c]{@{}c@{}}Collective\\ Name\end{tabular} & \begin{tabular}[c]{@{}c@{}}\# Occurrences\\ in Database\end{tabular} \\ \hline
MPI\_Allgather                                            & 2145                                                                 & MPI\_Gatherv                                              & 868                                                                  \\ \hline
MPI\_Allreduce                                            & 13576                                                                & MPI\_Reduce                                               & 11703                                                                \\ \hline
MPI\_Alltoallv                                            & 1287                                                                 & MPI\_Scatter                                              & 1252                                                                 \\ \hline
MPI\_Barrier                                              & 32954                                                                & MPI\_Scatterv                                             & 477                                                                  \\ \hline
MPI\_Gather                                               & 2630                                                                 & MPI\_Bcast                                                & 20417                                                                \\ \hline
\end{tabular}
\end{table}

\subsection{Collectives Pattern}
Based on our experimental study, we provide the following definitions. The following classifications are the fundamental building blocks of complex collective that we identified.

\begin{definition}[Complex or Fused Collective] A group $\mathcal{C} $ of more than one MPI collective.
\end{definition}

\begin{definition}[\textit{($\epsilon$, $\delta$)}-Repeated Collective]
% A multiple of complex collective $\mathcal{C}$ within an $\epsilon$ number of lines with $\delta$ occurrences.
A complex collective with a group $\mathcal{C}$ of more than one MPI collective within an $\epsilon$ number of lines with $\delta$ occurrences.
\end{definition}

\begin{definition}[Homogeneous Collective]
A complex collective $\mathcal{C}$ is \textit{homogeneous} if there exists only a unique type of collectives.  
\end{definition}

\begin{definition}[Mixed Collective]
A complex collective $\mathcal{C}$ is mixed if it is not a homogeneous complex collective.
\end{definition}

To make the above definitions clear, we give two examples according to our populated database. Example 1 shows a homogeneous $(5,2)$-repeated collective and Example 2 shows a mixed $(30,4)$-repeated collective (two homogeneous collectives sandwiched together).

\begin{verbatim}[Example 1]
  MPI_Allreduce() | line 93
  MPI_Allreduce() | line 98 
\end{verbatim}

\begin{verbatim}[Example 2]
 MPI_Allreduce() | line 200
 MPI_Allreduce() | line 217
 MPI_Allgather() | line 227
 MPI_Allgather() | line 230
\end{verbatim}

\subsection{All-Pair Search}
In this work, we implement a pair-wise search algorithm for our analysis. We sampled several potential MPI collectives combinations that usually occur together. Namely, the commonly used pairs such as: (\texttt{Gather}, \texttt{Scatter}), (\texttt{Allreduce}, \texttt{Allgather}), (\texttt{Allreduce}, \texttt{Alltoall}), (\texttt{Reduce}, \texttt{Bcast}), (\texttt{Gatherv}, \texttt{Gather}), (\texttt{Scatterv}, \texttt{Scatter}). We query each file with selected combinations of MPI collectives.

\begin{figure}[h].
  \centering
  \includegraphics[width=\linewidth]{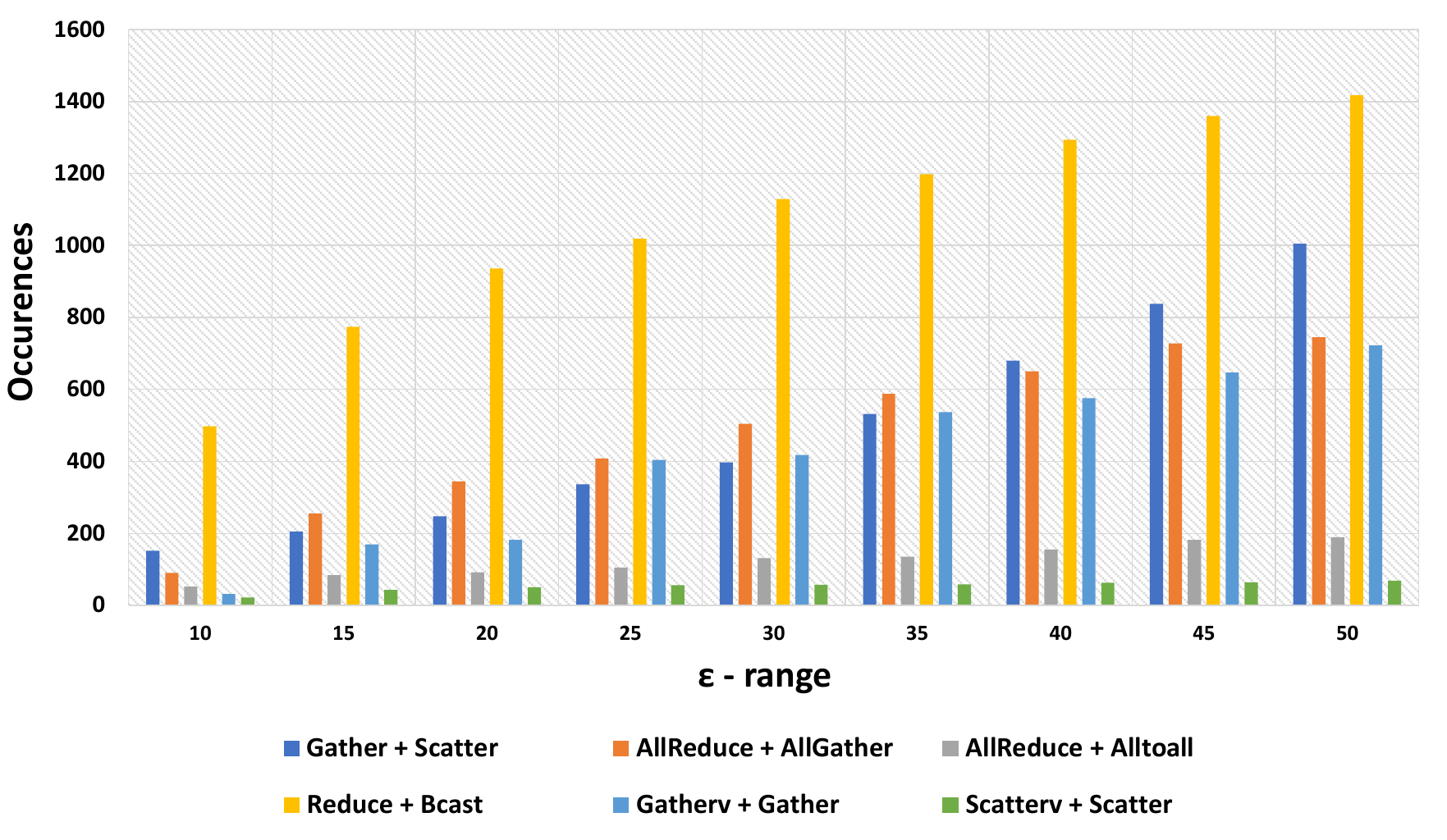}
  \vspace{-0.2truein}
  \caption{All-Pair Search Experiment for A Sample of Complex Collectives}
    \label{fig:result1}
\end{figure}

For our experiment, we set to find 
% $\delta$ number of occurrences of 
repeated collectives within some $\epsilon$ range. The detection method searches all combinations within the same file of the same repository and counts the aforementioned pairs that arise. The result of this experiment is depicted in Figure \ref{fig:result1}. The horizontal axis shows the $\epsilon$ range and the vertical axis shows the total number of occurrences for each sampled pair in our database. 

As it can be seen from the figure, as $\epsilon$ range is increased number of occurrences becomes higher. As it is evident from the figure, (\texttt{Reduce}, \texttt{Bcast}) pair is dominant among all of the other sampled pairs. We note that (\texttt{Gather}, \texttt{Scatter}) and (\texttt{Allreduce}, \texttt{Allgather}) are the next two dominant pairs that have the highest number of occurrences. To give some insight on what is the ratio of number of occurrences in Figure \ref{fig:result1} to total number of occurrences in the database (Table \ref{tab:db}), we give an example for (\texttt{Gather}, \texttt{Scatter}) with $\epsilon$ range as 50. This complex collective is used as much as 38\% and 80\% of total number of occurrences for \texttt{MPI\_Gather} and \texttt{MPI\_Scatter}, respectively, which shows the high possibility of fusion of these two collectives.

\subsection{Homogeneity Distribution}
\begin{figure}
    \centering
    \includegraphics[width=\linewidth]{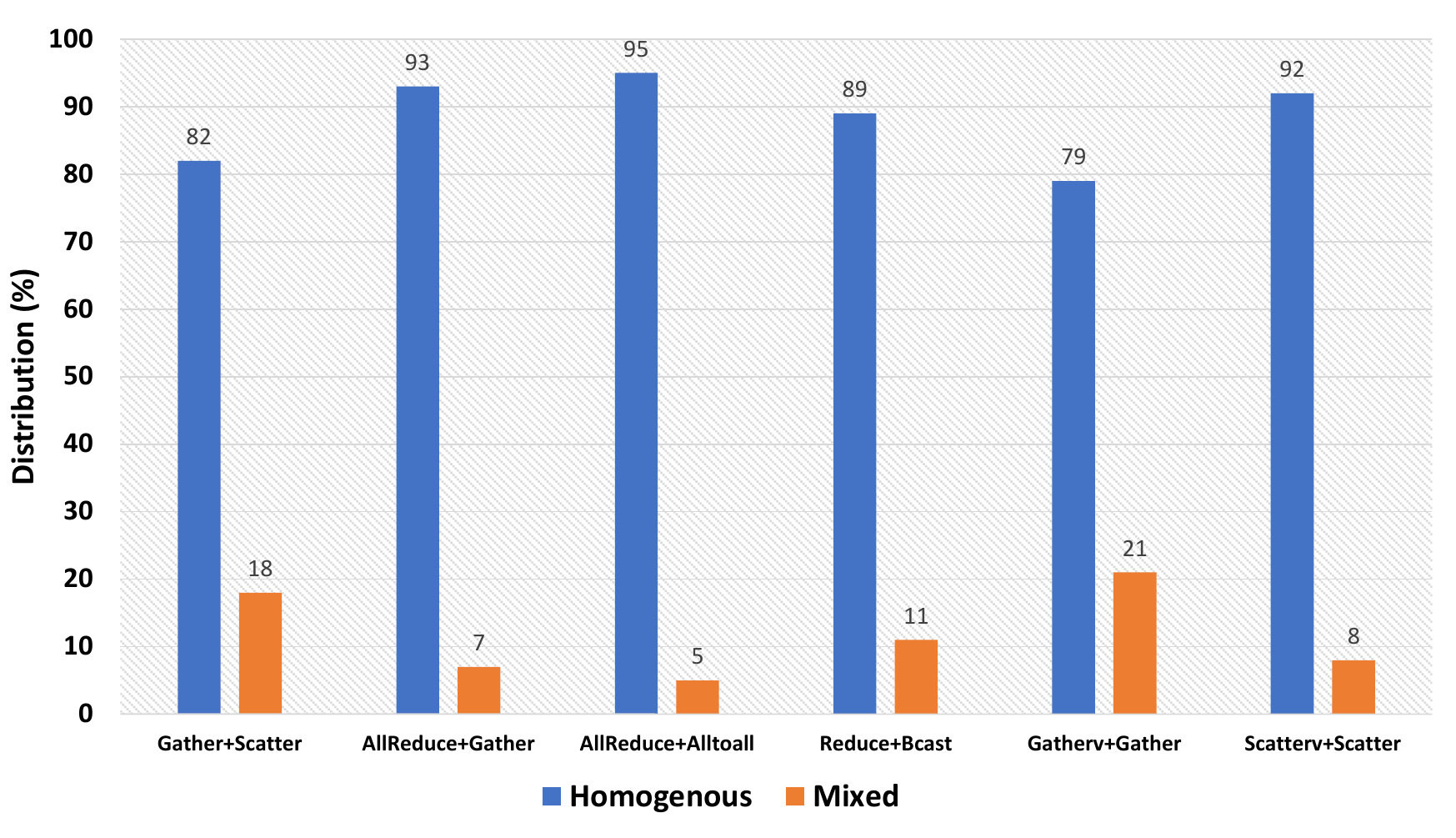}
    \vspace{-0.2truein}
    \caption{Homogeneity test: Y-axis shows the distribution of homogeneous versus mixed collectives for each sampled complex collective.}
    \label{fig:homogenity}
\end{figure}

In this experiment, we provide analysis for the distribution of homogeneous and mixed collectives. In this test, we focus on the correlation of individual collectives in each pair; we do not consider $\epsilon$ range.
% We relax $\epsilon$ range by not constraining it to a limit.
Although it is likely that there exists other combination of MPI collectives in between the two pairs, our study only focuses on the pair-wise pattern we described above. We recorded the number of occurrences found in each pair and tested their \textit{homogeneity} in Figure \ref{fig:homogenity}. Blue (red) bars represent homogeneous (mixed) collectives and Y-axis represent the ratio in percent. According to the distribution in this figure, homogeneous collectives outweigh mixed collectives significantly. (\texttt{Gatherv}, \texttt{Gather}) has the highest correlation while (\texttt{Allreduce}, \texttt{Alltoall}) has the lowest.
% Homogeneity test indicates the correlation between 

\section{Related Work}
The authors in \cite{Chunduri18} collected MPI usage profiles for around 100K jobs over a two-year period using a lightweight profiling tool, called Autoperf, that profiles and logs summarized statistics of MPI usage for production applications. Laguna et al. \cite{Laguna2019} presents a comprehensive study of MPI usage in applications at a significant scale with more than one hundred distinct MPI programs covering a large space of the population of MPI applications. In their study, they focus on understanding the characteristics of MPI usage with respect to the most used features, code complexity, and programming models and languages. The authors in \cite{Sultana2021} explore the usage of MPI for only fourteen mini applications within Exascale Computing Project (ECP) suite. Finally, the work \cite{Bernholdt2020} provides a summary of a survey conducted within the ECP community to gain information about how the MPI standard is currently used and how the various ECP projects are planning on using it to achieve exascale; but the scope of this paper is beyond current existing MPI usage as we provide insight on new MPI collective usage. 

\section{Discussion}
While MPI-Recon provides a straightforward yet efficient framework to mine MPI code bases there are still certain limitations. For instance, in this work, each line is treated equally with other lines. This might not be true in some cases; a line with simple addition is computationally lighter than a line with a convolution filter. Also, there might be one or more function calls within the complex collective that we did not take into account. This can change the analysis of potential complex collectives. Moreover, currently, the framework does not consider the collectives that are called multiple times (e.g., loops). Finally, some applications use their own MPI calls which is not currently supported in our work.

\section{Conclusion and Future Direction}
In this work, we propose a fully-automated pipeline to mine and characterize a large-scale collection of MPI code sources hosted in GitHub to explore new collective constructs. To efficiently store the information and perform complex analysis we use a database management system. Our pipeline is highly scalable and one can easily increase the size of the search space with enough storage space. We also believe that there exists many more complex collectives in the space for future studies.

We wish to extend our framework to search for other types of MPI communication routines including one-sided communication (RMA), point-to-point communication, neighborhood collectives, and many others. To account for loops and collectives that are called for multiple times, we plan to add a BSP detector plugin to our pipeline. A challenge in identifying the MPI collectives is the abstraction offered from other libraries such as \textit{HYPRE} \footnote{\url{https://computing.llnl.gov/projects/hypre-scalable-linear-solvers-multigrid-methods}} or \textit{Boost} \footnote{\url{https://www.boost.org/doc/libs/1_77_0/doc/html/mpi.html}}. A possible method to resolve the issue is to maintain a dictionary for translating the MPI collectives in the respective libraries. 
% Additionally, a random probing approach can be implemented to arbitrarily search for more interesting patterns. 
Additionally we seek to find other interesting patterns (\textit{e.g.,} tuples) and new probing methods (\textit{e.g.,} random probing).
% Notes:
% future work: RMA and ptpt, CUDA, handle wrappers, exclude boost, a flexible number of lists 
% report the number of repos
% 1) we might have overlap (we have searched for Gather but we got allreduce or even sth that we havent seached like Iallreduce)
% 2) We can have more complex analysis (different files but the same repo but still be fused)

% Searching is itself challenging, We can have 2 modes: one is pair-wise search another is just a query from the given MPI collective to just give us the meta-information

\section*{Acknowledgements}
% This work was supported, in part, by the NSF through award CCF-1919130; and by the NIH through award R44GM128533.
This work was supported, in part, by the NSF through awards CCF-1919130, CNS-1925504, and CCF-2151021, and by a grant from Red Hat. 
% and by AMD and Intel both through donated FPGAs, tools, and IP.

% \IEEEtriggeratref{16}
\bibliographystyle{IEEEtran}
\bibliography{sample-base,caad_refs_230414}

\end{document}